\documentclass{ws-ijmpd}

\newcommand{\beq}{\begin{equation}}
\newcommand{\eeq}{\end{equation}}
\newcommand{\bed}{\begin{displaymath}}
\newcommand{\eed}{\end{displaymath}}
\def\bea{\begin{eqnarray}}
\def\eea{\end{eqnarray}}

\newcommand{\nablasl}{{\slash \negthinspace \negthinspace \negthinspace \negthinspace  \nabla}}
\newcommand{\om}{\omega}

\begin{document}

\markboth{Owen Pavel Fern\'{a}ndez Piedra et. al}
{Object picture, quasinormal modes and late time tails of fermion perturbations
in stringy black hole with U(1) charges}

%
\catchline{}{}{}{}{}
%

\title{OBJECT PICTURE, QUASINORMAL MODES AND LATE TIME TAILS OF FERMION PERTURBATIONS IN STRINGY BLACK HOLES WITH U(1) CHARGES}

\author{OWEN PAVEL FERN\'{A}NDEZ PIEDRA}

\address{Grupo de Estudios Avanzados, Departamento de
F\'{i}sica, Facultad de Ingenier\'{i}a, Universidad de
Cienfuegos, Carretera a Rodas,km 4, Cuatro Caminos, Cienfuegos, Cuba \\opavel@ucf.edu.cu}

\author{FIDEL SOSA NU\~{N}EZ, JOSE BERNAL CASTILLO, YULIER JIMENEZ SANTANA}

\address{Grupo de Estudios Avanzados, Departamento de
F\'{i}sica, Facultad de Ingenier\'{i}a, Universidad de
Cienfuegos, Carretera a Rodas,km 4, Cuatro Caminos, Cienfuegos, Cuba}

\maketitle


\begin{abstract}
The aim of the present report is the study of massless fermion perturbations outside five-dimensional
stringy black holes with U(1) charges. The Dirac equation was numerically solved to obtain the
 time profiles for the evolving fermion fields, and the quasinormal
frequencies at intermediate times are computed by numerical Prony fitting
and semianalytical WKB expansion at sixth order. We also computed numerically
 the late-time power law decay factors, showing that there are in correspondence
 with previously reported results for the case of boson fields in higher dimensional
 odd space-times. The dependence of quasinormal frequencies with U(1) compactification
 charges are studied.
\end{abstract}


\section{Introduction}

The study of test field perturbations in a black hole background has been an interesting subject of research since the pioneer work
of Regge and Wheeler \cite{Regge:1957td,gpp,shutz-will,iyer-will,konoplya1,konoplya2,WKB6papers,zhidenkothesis,berti3}. Determining the time evolution of the perturbation outside the black hole we can obtain information about the particular properties of the background space-time geometry, as well as the stability of the compact object under such fluctuations and the form in which the field relaxes at very late times. It is known that, in asymptotically flat or de Sitter backgrounds, after an initial outburst of radiation, the evolution of massless perturbations is governed by quasinormal ringing, followed by power law or exponential tails at very late times. Such tails in general do not appear if the space-time geometry is asymptotically AdS, for wich the quasinormal ringing usually dominates the time evolution of the perturbations \cite{zhidenkothesis}.

In nature there are not isolated black holes, and for this reason the study of boson or fermion test fields around such compact objects is always motivated. Fermions describe some matter fields as those associated with neutrinos and gravitini, must which are essential for the study of black hole solutions coming from string/M theories. In this sense, a carefull analysis
of the quasinormal spectrum of these solutions provides a way to fix the parameters of the black hole and consequently of string/M theories \cite{cvetic1,cvetic2,cvetic3,horowitz1}.
 Also fermion perturbations of the gravitational bulk field equations for asymptotically AdS string/M theory black hole solutions turn out to be fundamental matter fields in the boundary conformal
field theory, providing an important class of objects relevant for condensed matter physics in a frame of the AdS/CFT correspondence \cite{Son:2007vk,Gubser:2009md,Hartnoll:2009sz}.

We take into account the first of the above facts as motivations to this paper, in which we consider the evolution of a spin-$1/2$ massless Dirac field in
the space-time of a black hole solution in five dimensions with three U(1) charges \cite{horowitz1}. This solution can also be obtained, along general
lines, considering intersection of branes and strings \cite{cvetic1,cvetic2,cvetic3}, and in this sense it describes a family of compact objects coming
 from string/M theory.

The structure of the paper is the following. After the introduction, in Section 2 we presents the general line element
 considered in the paper and write the fundamental equations to use for a study of Dirac perturbations in five dimensional space-times, obtaining a general
expression for the effective potential that describes the test field propagation. Section 3 is devoted to the numerical study of the evolution of the considered perturbation, as well as to present numerical and semi-analytical sixt order WKB results obtained for the quasinormal frequencies and its dependence on the charges
 of the black hole background, including the limit of large angular multipole numbers. In section 4 we present the numerical results concerning the relaxation
 of the perturbation at very late times, and propose an analytical form for the decay factors. Section 5 is devoted to the conclusions.

\section{Fundamental equations}

In five dimensions, there exist a family of black hole solutions parametrized by three independent charges, that can be obtained by the intersection of three 2-branes at a point, or from the intersection of a 2-brane and a 5-brane
with a boost, i.e, with a momentum along the common string \cite{cvetic3,horowitz1}. Upon toroidal compactification, the metric reads in both cases as:
\begin{equation}\label{metric}
ds^{2}=-A(r)dt^{2} + B(r) dr^{2} + C(r)d\Omega_{3}^{2}
\end{equation}
where $d\Omega_{3}^{2}$ denotes the line element of the round $3$-sphere and
 \begin{eqnarray}
 \nonumber
   A(r) &=& f(r)^{-2/3}\left(1-\frac{r_{H}^{2}}{r^{2}}\right) \\
   B(r) &=& f(r)^{1/3}\left(1-\frac{r_{H}^{2}}{r^{2}}\right)^{-1} \\ \nonumber
   C(r) &=& r^2f(r)^{1/3}
 \end{eqnarray}
where the function $f(r)$ is defined as
\begin{equation}\label{f}
f(r)=(1+\frac{r_{H}^{2}Q_{1}}{r^{2}})(1+\frac{r_{H}^{2}Q_{2}}{r^{2}})(1+\frac{r_{H}^{2}Q_{3}}{r^{2}}).
\end{equation}
In the above solution if at least one of the charges $Q_i$ is zero, the hipersurface $r=0$ is space-time singularity covered by the event horizon of the five dimensional black hole located at $r_H$. Moreover the case in which all the charges are non-zero corresponds to a regular black hole with an event horizon at $r=r_H$ and an inner horizon at $r=0$.

The evolution of a masless spin-$\frac{1}{2}$ fermion field in a five dimensional curved background is described by the Dirac equation:
\begin{equation}\label{}
   \nablasl \Psi=0
\end{equation}
where $\nablasl=\Gamma^{\mu}\nabla_{\mu}$ is the Dirac operator that acts on the five-spinor $\Psi$, $\Gamma_{\mu}$ are the curved space Gamma matrices, and the covariant derivative is
defined as $\nabla_{\mu}=\partial_{\mu}-\frac{1}{4}\om_{\mu}^{a b}\gamma_{a}\gamma_{b}$, with $\mu$ and $a$ being tangent and space-time indices respectively, related by the basis of orthonormal one forms
$\vec{e}^{\ a}\equiv e_{\mu}^{a}$. The associated conection one-forms $\om_{\mu}^{a b}\equiv\om^{a b}$ obey $d\vec{e}^{\ a}+\om^{a}_{\ b}\wedge\vec{e}^{\ b}=0$, and $\gamma^{a}$ are flat space-time gamma matrices
related with curved-space ones by $\Gamma^{\mu}=e^{\mu}_{\ a}\gamma^{a}$. They form a Clifford algebra
in five dimensions, i.e, they satisfy the anti-conmutation relations ${\gamma^{a},\gamma^{b}}=-2 \eta^{ab}$, with $\eta^{00}=-1$.

\par Under a conformal transformation of the metric in the form :
\begin{equation}
  g_{\mu\nu}=\Omega^{2}\tilde{g}_{\mu\nu} ,
\end{equation}
the five-spinor $\psi$ and the Dirac operator transforms as \cite{Gibbons,gibbons-rogatko}:
\begin{equation}
  \psi =\Omega^{-2}\tilde{\psi},
\end{equation}
and
\begin{equation}
 \nablasl \psi =\Omega^{-3} \ \tilde{\nablasl}\tilde{\psi},
\end{equation}
For a line element in the form $ds^{2}=ds_{1}^{2}+ds_{2}^{2}$, where $ds_{1}^{2} =g_{ab} (x) dx^a dx ^b$ and $ds_{2}= g_{mn}(y) dy^m dy ^n$, the Dirac operator $\nablasl$ satisfies the direct sum decomposition
\beq
\nablasl = \nablasl_x + \nablasl_y.
\eeq
 Thus, for two conformally related metrics, the validity of massless Dirac equation in one implies the validity of the same equation in the other. We use this fact to solve the Dirac equation in the curved space with the line element (\ref{metric}) by performing succesive conformal transformations that isolate the metric components that depend of the angular variables, and applied each time the direct sum decomposition of Dirac operator, until to obtain an equivalent problem in a spacetime of the form $M^{2}\times\sum^{3}$, where $M^{2}$ is two-dimensional Minkowsky spacetime in ($t,r_*$) coordinates (where $r_*$ is the tortoise coordinate defined by $dr_*=\sqrt{\frac{B}{A}}dr$) and $\sum^{3}$ is the metric describing the 3-sphere, in which the spectrum of massless Dirac operator is known. This procedure is general and has been applied previously to four dimensional stringly black hole by one of the authors \cite{owencqg2}.

The above method allow us to obtain, for each component of a Dirac spinor in the manifold  $M^{2}$ defined as:
\begin{equation}
\tilde{\varphi}_{\ell}^{(+)} = \left(
\begin{array}{c}
i\zeta_{\ell}(t,r) \\ \chi_{\ell}(t,r)
\end{array}
\right) ,
\end{equation}
the following equations:
\begin{equation}
i\frac{\partial\zeta_{\ell}}{\partial t}+\frac{\partial\chi_{\ell}}{\partial r_{*}}+\Lambda_{\ell}\chi_{\ell}=0
\end{equation}
and
\begin{equation}
i\frac{\partial\chi_{\ell}}{\partial t}-\frac{\partial\zeta_{\ell}}{\partial r_{*}}+\Lambda_{\ell}\zeta_{\ell}=0
\end{equation}
where
\beq \label{lambda}
\Lambda_{\ell}(r)= \sqrt{\frac{A}{C}}\left( \ell + \frac{3}{2} \right)
\eeq
This equations can be separated to obtain:
\begin{equation}
\frac{\partial^{2}\zeta_{\ell}}{\partial t^{2}}-\frac{\partial^{2}\zeta_{\ell}}{\partial r_{*}^{2}}+ V_{+}(r)\zeta_{\ell}=0
\label{tevol1}
\end{equation}
and
\begin{equation}
\frac{\partial^{2}\chi_{\ell}}{\partial t^{2}}-\frac{\partial^{2}\chi_{\ell}}{\partial r_{*}^{2}}+ V_{-}(r)\chi_{\ell}=0
\end{equation}
where:
\begin{equation}
V_{\pm}=\pm\frac{d\Lambda_{\ell}}{dr_{*}}+\Lambda_{\ell}^{2} .
\label{pot}
\end{equation}
The above equations gives the temporal evolution of Dirac perturbations outside the black hole spacetime \cite{owencqg2,splitfermion}. As the potentials $V_{+}$ and $V_{-}$ are supersymmetric to each other in the sense considered by Chandrasekhar in \cite{chandra}, then $\zeta_{\ell}(t,r)$ and $\chi_{\ell}(t,r)$ will develop similar time evolutions and
  will have the same spectra, both for scattering and quasi-normal. At this point it should be stressed that for the spinor $\tilde{\varphi}_{\ell}^{(-)}$, we have these two potentials again. In the following we will work with equation (\ref{tevol1}) and we
eliminate the subscript $(+)$ for the effective potential, defining $V(r)\equiv V_{+}(r)$.

\begin{figure}[htb!]
\begin{center}
\includegraphics[width=3.2in]{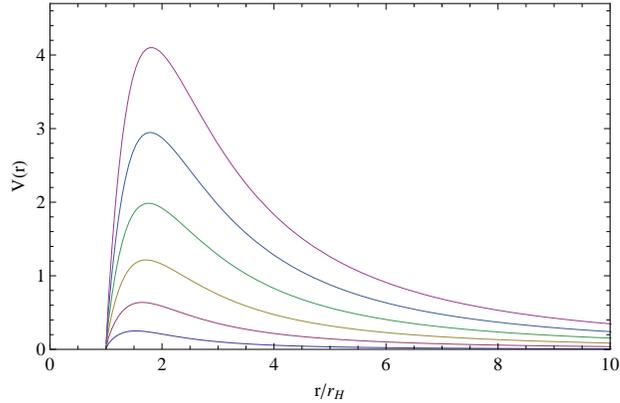}

\caption{\it Effective potencial for a (4+1)-stringy black hole with $\ell$ from 0(bottom) to 5(top) and $Q_1=Q_2=Q_3=1$.}
\label{potencial}
\end{center}
\end{figure}

For the stringy black hole solution considered in this report we have for the effective potencial $V(r)$ the following expression:
\begin{equation}
 V(r)=\frac{\lambda_\ell f^{-1}}{r^2}\left( 1-\frac{r_H^2}{r^2} \right) \left[1-\left( 1-\frac{r_H^2}{r^2} \right)^{\frac{3}{2}} \left[1-\frac{rf'}{2f}\left({1+\frac{r_H^2}{r^2}} \right)^{-1} \right]\right]
\label{pot1}
\end{equation}
where $\lambda_\ell = \ell+\frac{3}{2} $.

Figure (\ref{potencial}) shows the effective potential for different multipole numbers $\ell$ in the case of stringy black holes with  $Q_1=Q_2=Q_3=1$. The form of the effective potential is similar for other values of compactification charges, and as we can see, this assures the stability of the solution under fermion perturbations, due to its definite positive character.

\section{Time evolution of Dirac perturbations and quasinormal modes}
To integrate the equation (\ref{tevol1}) numerically we use the technique developed by Gundlach, Price and Pullin \cite{gpp}.

\begin{figure}[htb!]
\begin{center}

\includegraphics[width=3.2in]{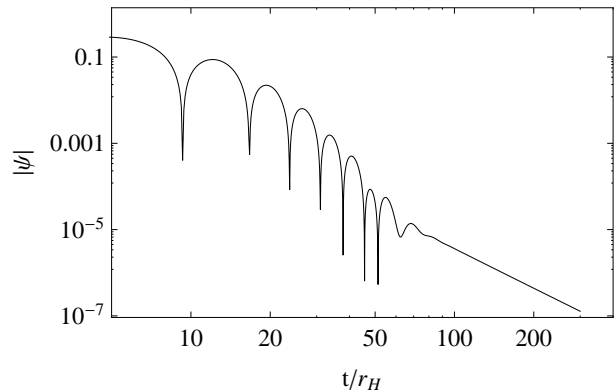}
\caption{\it Logaritmic plots of the time-domain evolution of $\ell=0$ and $Q_1=Q_2=Q_3=1$ massless Dirac perturbations.}
\label{perfil11}
\end{center}
\end{figure}

\begin{figure}[htb!]
\begin{center}

\includegraphics[width=3.2in]{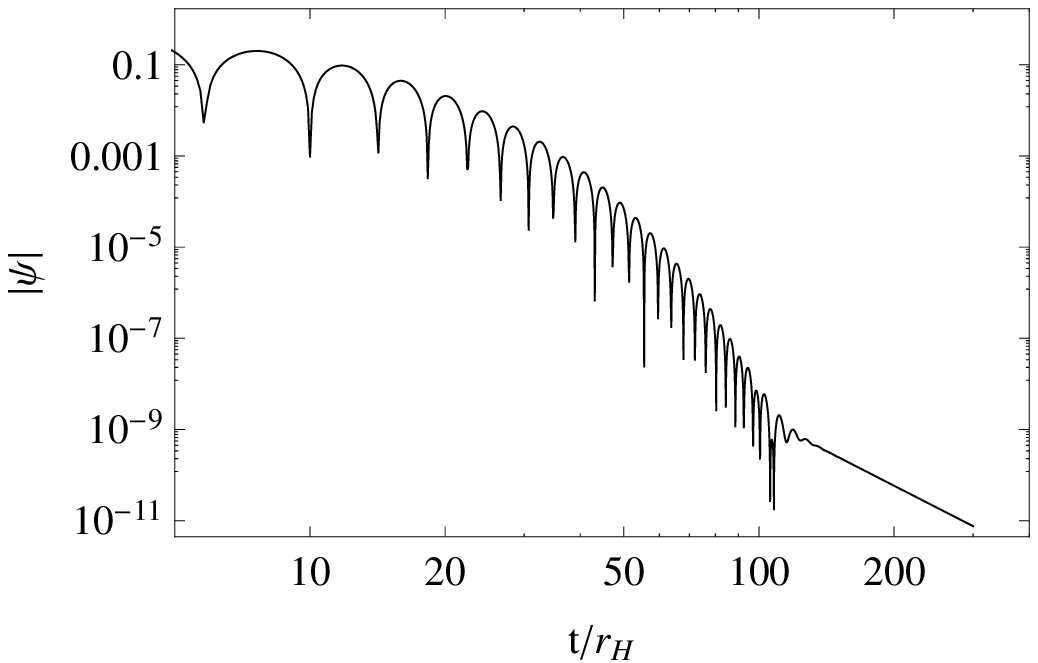}
\caption{\it Logaritmic plots of the time-domain evolution of $\ell=1$ and $Q_1=Q_2=Q_3=1$ massless Dirac perturbations.}
\label{perfil21}
\end{center}
\end{figure}

\begin{figure}[htb!]
\begin{center}

\includegraphics[width=3.2in]{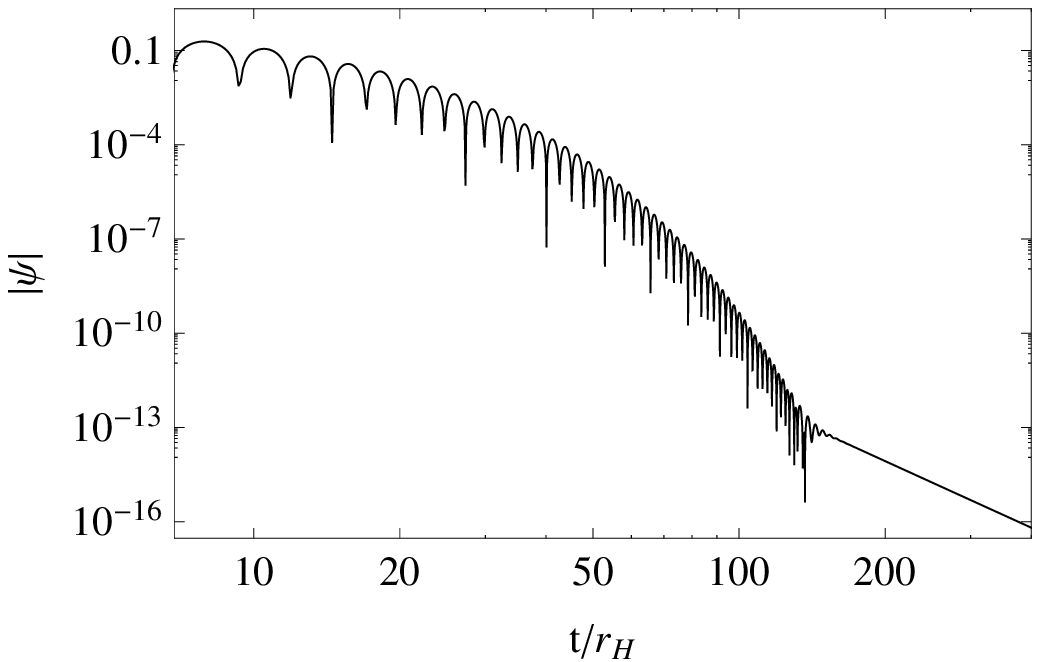}
\caption{\it Logaritmic plots of the time-domain evolution of $\ell=2$ and $Q_1=0$, $Q_2=Q_3=1$ massless Dirac perturbations.}
\label{perfil31}
\end{center}
\end{figure}

\begin{figure}[htb!]
\begin{center}
\includegraphics[width=3.2in]{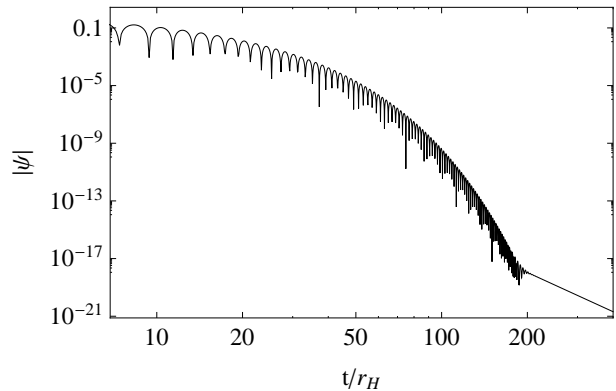}
\caption{\it Logaritmic plots of the time-domain evolution of $\ell=3$ and $Q_1=0$, $Q_2=Q_3=1$ massless Dirac perturbations.}
\label{perfil41}
\end{center}
\end{figure}

The obtained results in the case of massless Dirac fields in $(4+1)$-dimensional stringy black hole background can be observed as the time-domain profiles  showed in Figures (\ref{perfil11}) to (\ref{perfil41}). In such profiles $r=3 r_{H}$  and  the time is measured in units of black hole event horizon.

As is easily seen, the time evolution of Dirac perturbations outside five dimensional stringy black holes follows the usual dynamics for fields in other black hole backgrounds. After a first transient stage strongly dependent on the initial conditions and the point where the wave profile is computed, we observe the characteristic exponential damping of the perturbations associate with the quasinormal ringing, followed by a so-called power law tails at asymptotically late times.

To compute the quasinormal frequencies that dominated at intermediate times, we assume for the function $\zeta_{\ell}(t,r)$ in equation (\ref{tevol1}) the time dependence:
\begin{eqnarray}
\ \ \ \ \ \ \ \  \zeta_{\ell}(t,r)=Z_{\ell}(r)\exp(-i\om_{\ell}t)
\end{eqnarray}
Then, the function $Z_{\ell}(r)$  satisfy the Schrodinger-type equation:
\begin{equation}\label{finaleq}
\frac{d^{2} Z_{\ell}}{d r_{*}^{2}}+\left[\om^{2}- V(r)\right] Z_{\ell}(r)=0
\end{equation}
The quasinormal modes are solutions of the wave equation (\ref{tevol1})
with the specific boundary conditions requiring pure out-going waves
at spatial infinity and pure in-coming waves on the event horizon.
Thus no waves come from infinity or the event horizon.

The quasinormal frequencies were computed using two different
methods. The first method uses directly the numerical data obtained previously, and fit this data by superposition of damping exponents. This numerical fitting scheme, known as Prony method, allow us to obtain very accurate results for the fundamental and first overtones \cite{berti3,zhidenkothesis}. For higher overtones is very difficult to be implemented, because we need to do a fitting with a great number of exponentials, and also we need to know very well the particular time in which quasinormal ringing begins, a difficult point to be solved in general.  For this reason in the following we only presents the quasinormal frequencies, using this method, for the first two overtone number.

The second method that we employed is a semi-analytical approach to solve equation
(\ref{finaleq}) with the required boundary conditions, based in a
WKB-type approximation, that can give accurate values of the lowest
( that is longer lived ) quasinormal frequencies, and  was used in
several papers for the determination of quasinormal frequencies in a
variety of systems \cite{shutz-will,iyer-will,konoplya1,konoplya2,WKB6papers,WKB6papers1,WKB6papers2,WKB6papers3,WKB6papers4,WKB6papers5,nuestro,splitfermion}.

\begin{table}[htb!]
 \tbl{\it Dirac quasinormal frequencies \(\omega r_{H}\) in
(4+1)-stringy black hole with Q1=0.5, Q2=1, Q3=1.8   for $\ell=0$ to
$\ell=4$. The frequencies are measured in units
of the black hole horizon radious $r_{H}$.}
  {\begin{tabular}{|c|c|c|c|}
      \hline
      \hline
       $\ell$ & $n$ & Sixth order WKB & Prony  \\
      \hline
      $0$ & $0$  & $0.4349 -0.1712i$    &  $ 0.4346 -0.1709i $ \\
      \hline
  $1$ & $0$ &  $0.7445-0.1795i$     & $ 0.7443-0.1792i $ \\
      \hline
  $2$ & $0$ &  $1.0518-0.1806 i$    &  $ 1.0518-0.1806 i $ \\
      \hline
 $2$ & $1$  & $ 1.0073-0.5489i$ & $ 1.0072-0.5488i $ \\
     \hline
 $3$ & $0$  & $ 1.3571-0.1809i$ & $ 1.3571-0.1809 $ \\
      \hline
$3$ & $1$   &  $ 1.3224-0.5469i$    & $ 1.3224-0.5469i $ \\
     \hline
$3$ & $2$   & $1.2540-0.9263i $ & - \\
      \hline
$4$ &  $0$  & $ 1.6623-0.1810i $    & $ 1.6623-0.1810i  $ \\
      \hline
 $4$ & $1$  & $1.6335-0.5458i$  & $1.6335-0.5458i$ \\
      \hline
$4$ & $2$   &  $1.5769-0.9196i$ & - \\
     \hline
$4$ & $3$   & $1.4942-1.3083i$  & - \\
      \hline
       \hline
   \end{tabular}\label{frecuenciasdirac1}}
 \end{table}

\begin{table}[htb!]
 \tbl{\it Dirac quasinormal frequencies \(\omega r_{H}\) in
(4+1)-stringy black hole with Q1=0, Q2=1, Q3=1 for $\ell=0$ to
$\ell=4$. The frequencies are measured in units
of the black hole horizon radious $r_{H}$.}
  {\begin{tabular}{|c|c|c|c|}
      \hline
      \hline
       $\ell$ & $n$ & Sixth order WKB & Prony  \\
      \hline
      $0$ & $0$  & $0.5101-0.2063i$ &  $ 0.5098-0.2059i $ \\
      \hline
  $1$ & $0$ &  $0.8702-0.2152i$     & $0.8700-0.2151i $ \\
      \hline
  $2$ & $0$ &  $1.2270-0.2161i$     &  $1.2270-0.2161i $ \\
      \hline
 $2$ & $1$  & $ 1.1789-0.6579i$ & $1.1788-0.6578i $ \\
     \hline
 $3$ & $0$  & $ 1.5829-0.2163i$ & $1.5829-0.2163i $ \\
      \hline
$3$ & $1$   &  $ 1.5445-0.6547i$    &$1.5445-0.6547i $ \\
     \hline
$3$ & $2$   & $1.4720-1.1106i $ & - \\
      \hline
$4$ &  $0$  & $ 1.9379-0.2163i $    &$  1.9379-0.2163i  $ \\
      \hline
 $4$ & $1$  & $1.9063-0.6530i$  & $1.9063-0.6530i$ \\
      \hline
$4$ & $2$   &  $1.8452-1.1014i$ &$ - $ \\
     \hline
$4$ & $3$   &$ 1.7590-1.5689i$  & $ - $ \\
      \hline
       \hline
   \end{tabular}\label{frecuenciasdirac1}}
\end{table}

\begin{figure}[htb!]
\begin{center}
\includegraphics[width=3.0in]{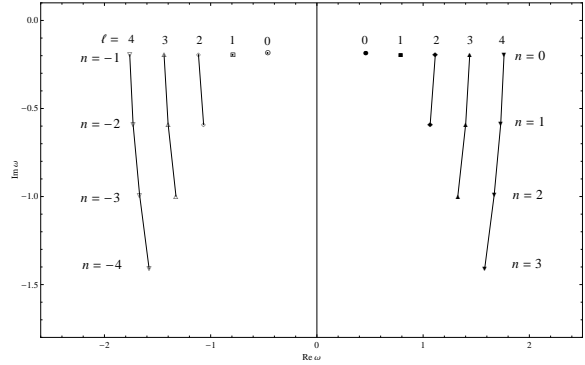}
\caption{\it{Massles Dirac quasinormal frequencies of (4+1)-stringy black holes with $Q_1=Q_2=Q_3=1$.}}
\label{QNM}
\end{center}
\end{figure}

\begin{figure}[htb!]
\begin{center}
\includegraphics[width=3.0in]{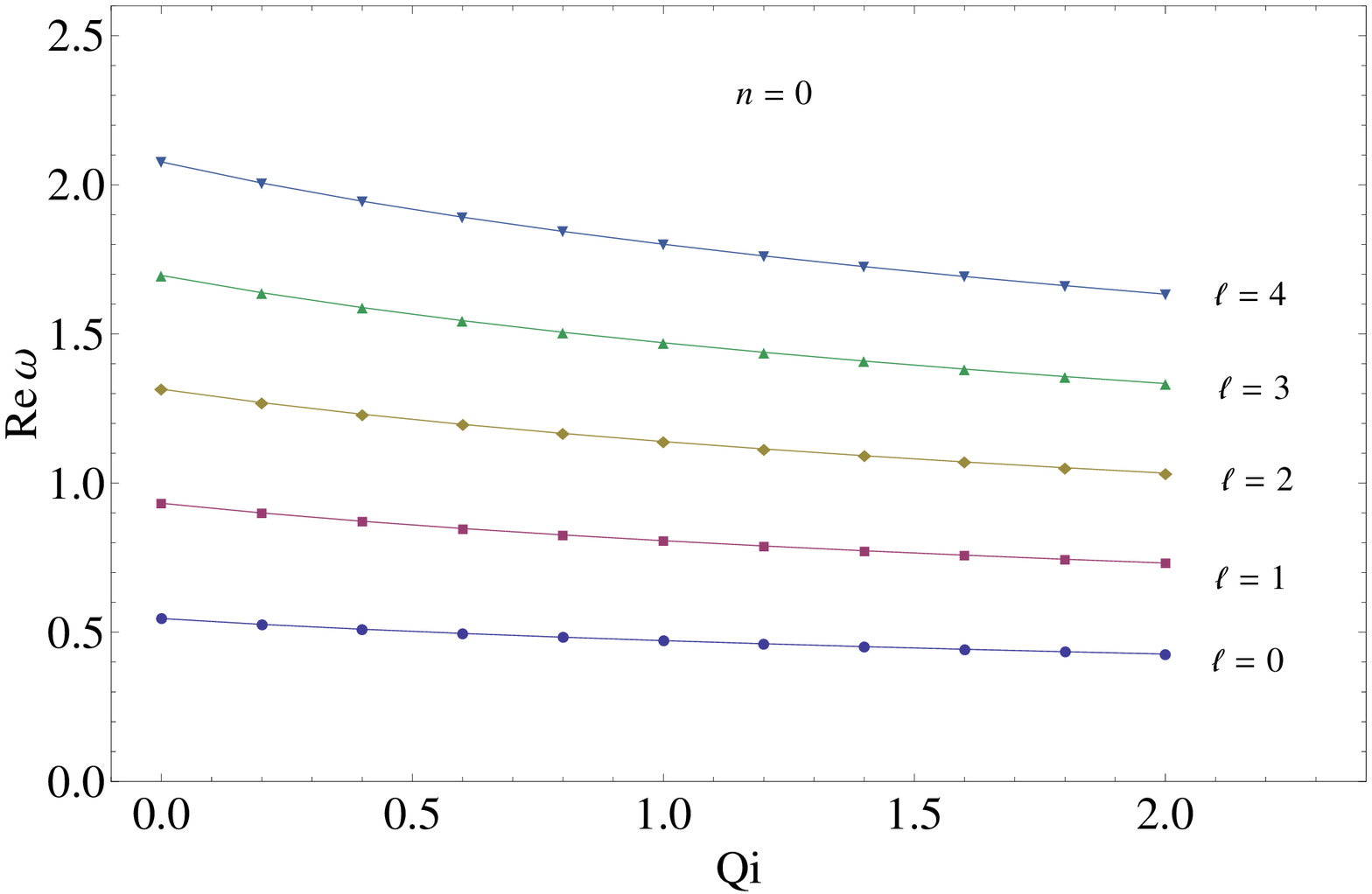}
\caption{Dependence upon charge $Q_i$, $i=1,2,3$ of the real part of the massless Dirac quasinormal frecuencies of stringly black hole with the other two charges fixed $Q_{j}=1$, $j\neq i$. The results correspond to the first overtone for multipolar number from $\ell=0$ to $\ell=4$.}
\label{o1}
\end{center}
\end{figure}

\begin{figure}[htb!]
\begin{center}
\includegraphics[width=3.0in]{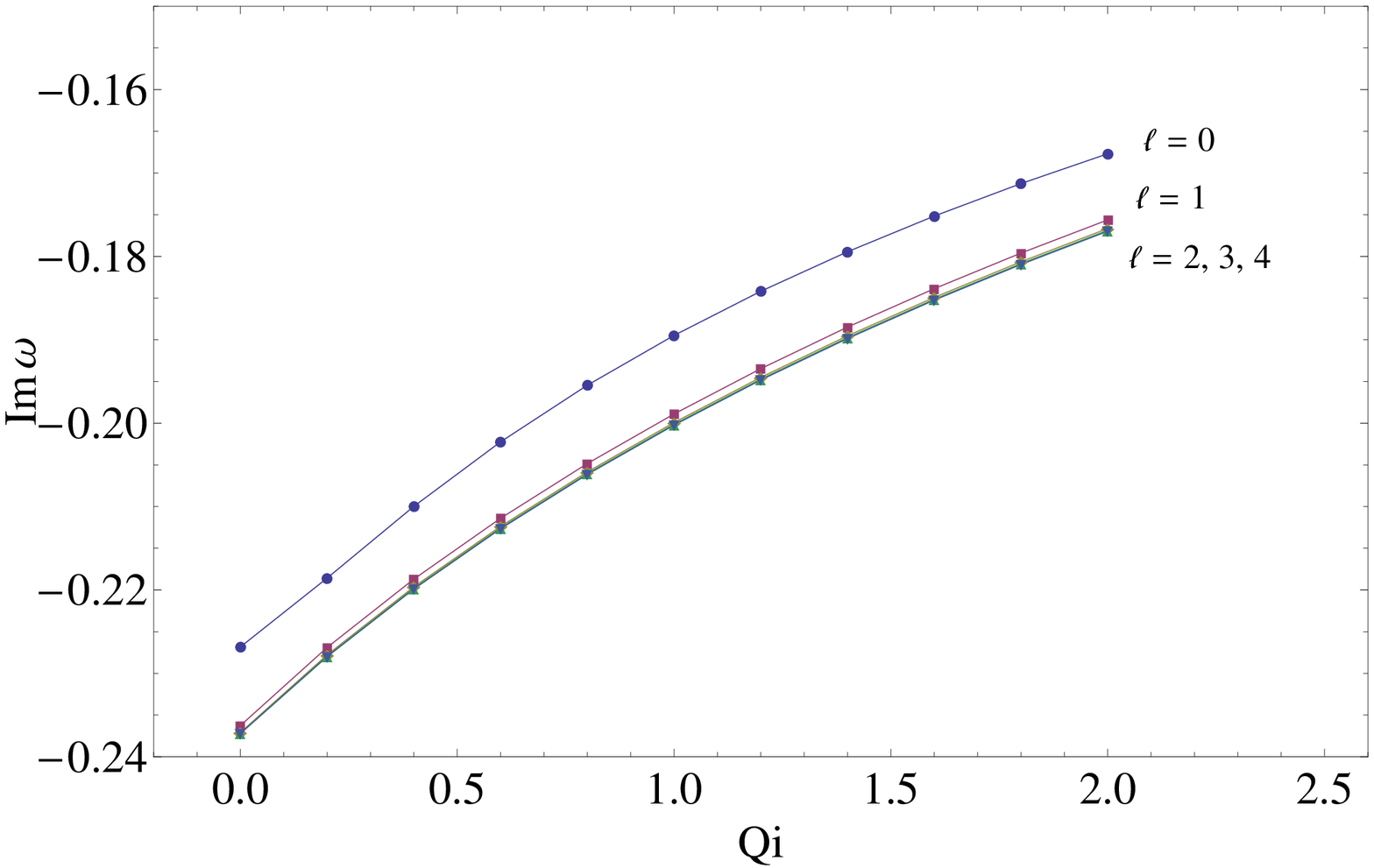}
\caption{Dependence upon charge $Q_i$, $i=1,2,3$ of the imaginary part of the massless Dirac quasinormal frecuencies of stringly black hole with the other two charges fixed $Q_{j}=1$, $j\neq i$. The results correspond to the first overtone for multipolar number from $\ell=0$ to $\ell=4$.}
\label{o2}
\end{center}
\end{figure}

Tables 1 and 2 presents the values obtained for the quasinormal frequencies with some multipole numbers $\ell$ and different set of charges. We also present the result for a particular stringy black holes with $Q_1=Q_2=Q_3=1$ in Figure (\ref{QNM}). As it is observed, the sixth order WKB approach gives results in agreement with those obtained by fitting the numerical integration data using Prony technique. As it is expected the oscillation frequency increases for higher multipole and fixed overtone numbers. Increasing $\ell$ the magnitude of the negative imaginary part of the fundamental overtone ($n=0$) increases while for higher overtones the opposite situation arises. For a fixed angular number $\ell$, the real part of the oscillation frequencies decreases as the overtone number increases, and the magnitude of the imaginary part increases. Then, modes with higher overtone numbers decay faster than low-lying ones. As we expected for stability, all quasinormal frequencies calculated in this work have a well defined negative imaginary part.

Figures (\ref{o1}) and (\ref{o2}) show the dependence of the quasinormal modes with compactification charges. As one of the charges increases, leaving fixed the other two, the real part and the absolute value of the imaginary part of the quasinormal frequencies decreases, but the rates of decreasing are different. Then, as the charges increases, the modes labeled by the same angular multipole numbers quickly becomes less damped and are long lived.

An interesting case arises for large angular multipole numbers because in this limit the first order WKB approximation becomes exact and we can obtain an analytical result. In this limit the effective potential (\ref{pot1}) can be written as
\begin{equation}
U(r)=\ell^{2}\Delta(r)
\label{pot2}
\end{equation}
where $\Delta(r)=\frac{\Xi(r)}{r^2 f(r)}$ and $\Xi(r)=1-\frac{r_H^2}{r^2}$. Then the first order WKB approximation gives for the quasinormal frequencies in this limit the result:
\begin{equation}\label{wkb_large_l}
\omega^{2}=\ell^{2}\Delta(r_{m})-i\ell(n+\frac{1}{2})\sqrt{-2\frac{d^{2}\Delta(r)}{dr_{*}^{2}}|_{r=r_{m}}},
\end{equation}
being $r_{m}$ the point in which the asymptotic effective potential (\ref{pot2}) reach its peak. This value can be determined as the maximum root of the equation
\begin{equation}
-\Xi(r)\left[2f(r)+r\frac{df(r)}{dr}\right]+rf(r)\frac{d\Xi(r)}{dr}=0.
\end{equation}
In the particular case of equal charges $Q_{1}=Q_{2}=Q_{3}=Q$ we obtain the result:
\begin{equation}
r_{m}=r_{H}\sqrt{1+Q+\sqrt{1+Q+Q^{2}}}
\label{root}
\end{equation}

\section{Late Time Tails}
Another important point to study is the relaxation of the perturbing fermion field outside the black hole at late times \cite{Price,burko}. It is a known result that in higher dimensional Schwarzschild black hole neutral massless boson fields had a late-time behavior dominated ( for a fixed $r$ and each multipole moment $\ell$ ) by a factor $t^{-(2\ell+D-2)}$ for odd $D$-dimensions and $t^{-(2\ell+3D-8)}$ for even $D$-dimensions  \cite{cardoso1}.

To study the late-time behavior, we numerically fit the profile data
obtained in the appropriate region of the time domain, to extract the power law exponents that describe the relaxation. As a test of our numerical fitting scheme, we obtained the power law exponents for the massless Dirac field considered in this paper in the space-time corresponding to higher dimensional Schwarzschild black hole. As we expected, the results obtained are consistent with the power law falloff mentioned in the previous paragraph.

\begin{figure}[h!]
\centering
\mbox{{\includegraphics[width=3.2in]{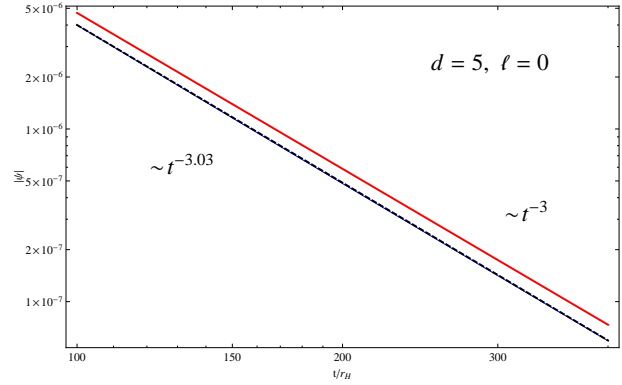}}}
\caption{{\it{Tail for $\ell=0$ and $Q_1=0$, $Q_2=Q_3=1$. The power-law coefficients were estimated from numerical data represented in the dotted line. The full red line is the possible analytical result.}}}
\label{tail1}
\end{figure}

\begin{figure}[h!]
\centering
\mbox{{\includegraphics[width=3.2in]{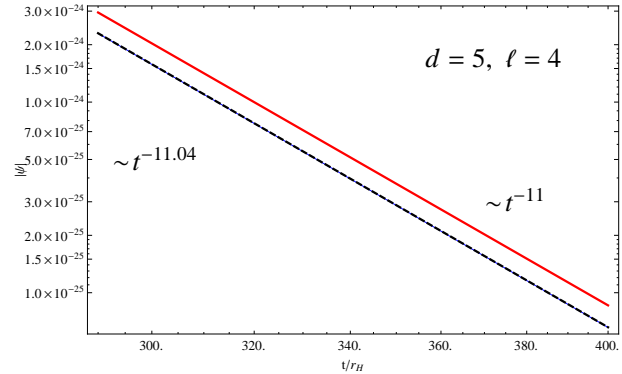}}}
\caption{{\it{Tail for $\ell=4$ and $Q_1=0$, $Q_2=Q_3=1$. The power-law coefficients were estimated from numerical data represented in the dotted line. The full red line is the possible analytical result.}}}
\label{tail2}
\end{figure}

Figures (\ref{tail1}) and (\ref{tail2}) show two representative examples of the results obtained, for a five dimensional stringy black hole with one charge zero and the other two $Q_{2}=Q_{3}=1$. In general, all our numerical results indicate that, in five dimensions, the decay in the Dirac case is governed by factors of the form  $t^{-(2\ell+3)}$. This result is in complete correspondence with the general result obtained for bosons fields in higher dimensional space times: if $D$ is the dimension of the space time then the late time tail is described by the power-low falloff $\varPsi\sim t^{-(2\ell+D-2)}$ for odd $D$ \cite{cardoso1}. It seems that the fermion or boson character of the test field have no influence in the late falloff of the perturbations also in higher dimensions. Then, we can conclude that, outside five dimensional stringy black holes, as well as Schwarzschild black hole, the massless Dirac field shows identical decay at late times.

However, we remark at this point that this dependance is only a result consistent with our numerical data. A simple analytical argument to support this late time behaviour do not exist, in contrast to the case for boson fields, in which the general form of the effective potential is suitable to expand for large values of the tortoise coordinate \cite{cardoso1} and then extract the above power law behavior directly from this asymptotic expansion. The problem related with the analytical determination of the decay factors for fermion perturbations in higher dimensional stringy black holes remains open.


\section{Concluding remarks}

In this report we considered the evolution of massless Dirac test field in the space time corresponding to five dimensional black hole solutions coming from intersecting branes in string theories.

After the initial transient epoch, the evolutions is dominated by quasinormal modes, and at late times by a power-low falloff. We computed the quasinormal frequencies for different values of compactification charges using two different approaches, 6th-order WKB formula and time domain integration with Prony fitting of the numerical data, obtaining by both methods very close numerical results. The results for the dependence of the quasinormal frequencies with charges appears to be universal for all values of this parameter.

We also computed the decay factors for the late time relaxation of fermion perturbations in five dimensional stringy black holes and show that the result is similar for those obtained for boson field in higher odd dimensional space-times.

It should be interesting to study the case of more higher dimensions, when remains open if the fermion decay picture at late times obey the same power law behaviour of boson fields. Another important question is the analytical study of this late time decay factors, taking into account that for potentials typical of fermion fields do not exist a simple analytical argument to approach this problem, as in the case of boson fields.

Stringy black holes obtained by intersection of branes are known for dimensions up to D=9, and then it would be interesting to study the evolution of test fields in this higher dimensional backgrounds. In future reports we will complete this studies to gain a more complete knowledge about the evolution of fermion as well as boson perturbations in this interesting physical systems.

\section*{Acknowledgments}

We are grateful to Professor Elcio Abdalla and Dr. Jeferson de Oliveira at IFUSP, Brazil, for useful discussions about quasinormal modes and Universidad de Cienfuegos for technical support.


\end{document}